\scrollmode
\documentstyle[12pt,psfig]{l-aa}
\begin{document}

\thesaurus{08    % A&A Section 8:  stars
              ( 
               08.05.3;  % stars:evolution,
               08.16.7 PSR B1744--24A %pulsars:individual
               .)} % Stars: structure of.

\title{The 
eclipsing binary millisecond pulsar PSR B1744--24A -- 
possible test for a magnetic braking mechanism}

\author{ Ene
Ergma\inst {1} and Marek J. Sarna\inst {2}} 
\offprints{Ergma}
\institute{Physics Department, Tartu University, \"Ulikooli 18, EE2400
       Tartu, Estonia \\
email ene@physic.ut.ee
\and
N. Copernicus Astronomical Center,
       Polish Academy of Sciences,
       ul. Bartycka 18, 00--716 Warsaw, Poland. \\
email sarna@camk.edu.pl}

\date{Received ;accepted}

\maketitle
\markboth{E. Ergma and M. J. Sarna: The eclipsing binary PSR B1744--24A..}{}

\begin{abstract}
As presented by Nice et al. (2000), long--term timing of the eclipsing 
binary PSR B1744--24A shows that the 
orbital period of this system  decreases with a time--scaleof only $\sim$ 
200 Myr. To explain the much faster orbital period decay than that predicted 
by only emission of the gravitational waves ($\sim$ 1000 Myr) we propose 
that the orbital evolution of this system is also driven 
by magnetic braking . If magnetic braking is to explain the rapid
decay of the orbit, then $\lambda$ characterizing the effectiveness of 
the dynamo action in the stellar convection zone in 
the magnetic stellar wind formula must be equal to 1. 
 
\end{abstract}


\section{Introduction}

The formation and evolution of a very low--mass binary system consisting of a 
neutron star (millisecond pulsar) and a 0.1 $M_\odot$ companion is not 
yet well understood. One of the main observational features of these systems 
is that they are eclipsing.
One interesting eclipsing binary system, PSR B1744--24A, is in the 
globular cluster Terzan 5. The neutron star is an 11.56 ms pulsar in a 
1.8 h orbit with a low--mass companion (Lyne et al. 1990). The  
duration of the eclipse in this system is very variable and is never less 
than one--third 
of the orbital period. Lyne et al.(1990) estimated that the energy flux 
from the pulsar at the companion surface (isotropic radiation is assumed) is 
$<$ 2$\times 10^{10}$ erg s$^{-1}$cm$^{-2}$ which is near  the critical 
value when the irradiation may influence the structure of a low--mass star 
(Podsiadlowski 1991). However, this value is at least a factor of 15 less 
than for another eclipsing millisecond binary pulsar system PSR B1957+20 
(Fruchter et al. 1990). This small irradiative flux value and great 
variations in duration of the eclipse lead Lyne et al. (1990) to the 
conclusion that in this system Roche lobe overflow may occur.

Recent  long--term timing observations of the eclipsing binary pulsar 
PSR B1744--24A at 
 the VLA  and Green Bank, show that  
the orbital period of PSR B1744--24A has decreased, with a time--scale of only
$P_{orb}/\dot{P}_{orb}$$\sim$200 Myr. This is five times faster  than if the
orbital period decay is driven only by the gravitational wave radiation 
(Nice et al. 2000).   

The key question of interest is: how did this system evolve? 

In this short note we  show that the new observational data for 
the orbital period decay of PSR B1744--24A may give us a unique 
opportunity to test the magnetic braking mechanism if the  observed 
orbital period decay is  secular.

\section{General picture of low--mass binary system evolution}

More than twenty years ago it was realized that to produce short  orbital 
period low--mass binary systems (cataclysmic variables and low--mass X--ray 
binaries -- LMXB) it is necessary to include two  mechanisms for 
orbital angular momentum losses: gravitational wave radiation and/or magnetic 
braking. 
Low--mass unevolved stars (main--sequence stars) will fill their Roche--lobe 
only for orbital periods less than 12 hours and the mass transfer due to 
gravitational wave radiation mechanism 
never exceeds $\sim 10^{-10}M_\odot$/yr. This produces an X--ray luminosity 
of $\sim 10^{36}$ erg/s. Observations, however, show that many LMXBs in 
this orbital period range have X--ray luminosities
that are one or two orders of magnitude higher.

To explain this discrepancy Verbunt \& Zwaan (1981) introduced additional 
orbital angular momentum loss by magnetic braking which was able to 
drive a much higher mass--transfer rate. In view of these much higher mass 
transfer rates the companion will be driven out of thermal equilibrium. 
However magnetic braking will not work forever. As soon as
the companion mass has decreased to M $\sim $ 0.3 $M_\odot$ (fully convective 
star) an orbital period near the 3 hours, the rate of angular momentum loss 
by magnetic braking is expected to 
either vanish or to drop considerably (Spruit \& Ritter 1983), causing the 
mass--loss
rate to drop, and the secondary star subsequently relaxes to its thermal 
equlibrium state. The stellar radius decreases and the star detaches  from 
its Roche lobe. During the 
detached phase, orbital angular momentum loss by gravitational wave radiation 
will continue, and when the orbital period has been reduced to 2 hours, the 
companion (of mass $\sim$0.3 $M_\odot$) again fills its Roche lobe. This 
is the so--called standard cataclysmic variable (CV) evolutionary scenario.  

In 1985 Tutukov et al. discussed  what will happen with 
a low--mass binary if the star filling its Roche lobe is a slightly evolved 
star (so--called ``turn--off main--sequence star''). In this 
scenario  the binary  evolves towards
very short orbital periods. Due to the chemical composition gradient, 
the secondary star does not become fully convective when its mass has 
reduced to 0.3$~M_\odot $ and  magnetic braking does not vanish. Between 
orbital periods of two to 
three hours, the mass transfer rate is very low, and apparently during this 
time the system may be in the ``propeller'' stage (Ergma \& Sarna 1996). 
According to the standard CV
evolutionary picture, the magnetic braking mechanism vanishes near an 
orbital period of 3 hours. However in the latter scenario, it will also 
work for smaller values of the orbital periods .

\section{Model}

The evolutionary sequences for the secondary star in a low--mass binary were 
computed using a standard one--dimensional stellar evolution code based 
on a Henyey--type code developed by Paczy\'nski (1970) and adapted to low--mass 
main--sequence stars. For more detail about the computer program see 
Muslimov \& Sarna (1993) and Sarna \& De Greve (1994, 1996).

We compute the loss of orbital angular momentum due to gravitational 
radiation using the formula presented by Landau \& Lifshitz (1971).

\begin{equation}
\frac{\dot{J}}{J}_{GWR}= -\frac{32}{5}(2\pi)^{\frac{8}{3}}\frac{G^{\frac{5}
{3}}}{c^5}\frac {M_1M_2}{M^{\frac{1}{3}}}P_{orb}^{-\frac{8}{3}}
\end{equation}
where $M_1$, $M_2$, $M$ and $P_{orb}$ 
are the primary mass, secondary mass, total mass and orbital period of the 
system.
We assume that the companion is being spun down by the magnetic braking 
and  its spin and orbital rotation are synchronized 
at the cost of orbital angular momentum loss. We adopt the standard 
formula (see Mestel 1968, Mestel \& Spruit 1987) for the rate of orbital 
angular momentum losses due to magnetic braking of the companion.

\begin{equation}
\frac{\dot{J}}{J}_{MB} = -5.0\times 10^{-29}(2\pi)^{\frac{10}{3}}G^{-\frac{2}{3}}
(\frac{k_2}{\lambda})^2\frac{M^{\frac{1}{3}}R_2^4}{M_1}P_{orb}^{-\frac{10}{3}}
\end{equation}

where $k_2$ is the radius of gyration of the secondary star, $k_2$ =0.1.
This formula contains a poorly known parameter $\lambda$ ($\sim$ 0.7--1.8) 
characterizing the effectiveness of the dynamo action in the stellar 
convection zone (Verbunt \& Zwaan 1981). 
To take account of the angular momentum loss that accompanies mass loss 
during the ``propeller phase'', we use a formula based on that used to 
calculate angular momentum loss via a stellar wind (Paczy\'nski, 1967, 
Sarna \& De Greve, 1994)

\begin{equation}
\frac{\dot{J}}{J}_{WIND} =f_1f_2 \frac{M_1\dot{M_2}}{M_2M},~~
\dot{M}=f_1\dot{M_2},~~ \dot{M_1}= -\dot{M_2}(1-f_1)
\end{equation}

where  $f_1=\frac{M_{ej}}{M_{acc}}$ and $\dot{M_2}$($\leq$0) is the mass--loss rate from the secondary star; $\dot{M_1}$($\geq$0) is the accretion rate on to the neutron star, $\dot{M}$($\leq$0) is the rate of the total mass loss from the system;
 $f_1$ is the ratio of the mass ejected by the neutron star to that 
accreted by the neutron star and $f_2$ is defined as the effectiveness of 
angular momentum loss during mass transfer (Sarna \& De Greve, 1994,1996).  

We consider the following model (similar scenarios have been discussed by 
Shaham \& Tavani 1991, Klu\'zniak et al. 1992 and Ergma \& Sarna 1996) 
for the PSR B1744--24A system. 
The progenitor of this system is a low--mass ($<$1--1.5$M_\odot$) star + old 
neutron star. The secondary star fills its Roche lobe, as a ``turn--off 
main--sequence star''. At the beginning, the mass transfer rate proceeds on 
a thermal time--scale and the neutron star will spin up to a millisecond 
period. After that, the mass transfer rate drops and the system enters the 
``propeller stage'' and the pulsar spins down. 

\begin{table}
\begin{center}
\begin{tabular}{cccc}
\multicolumn{4}{c}{Table 1}\\
Case & $\lambda$ & $\dot{P}_{orb}/10^{-12}$ & $\tau $\\
&  & [$s s^{-1}$] & [Myr]\\
(A) & 0.7 & 1.16 & 174 \\
(A) & 0.9 & 1.10 & 184 \\
(B) & 0.7 & 2.44 & 83 \\
(B) & 1.0 & 0.83 & 244 \\
(B) & 1.2 & 0.69 & 292 \\
(B) & 1.5 & 0.60 & 337 \\
(C) & 0.7 & 1.77 & 114 \\
(C) & 0.9 & 1.21 & 167 \\
\end{tabular}
\end{center}
\end{table}

As an example we calculated several evolutionary sequences: (I) $M_2$ = 
1$M_\odot$, $M_{1}$=1.4 $M_\odot$, $P_i$(RLOF)= 1 day;  (II) $M_2$ = 
1.5$M_\odot$, $M_{1}$=1.4 $M_\odot$ $P_i$(RLOF)= 1.02 days ($P_i$(RLOF) 
is initial orbital period when the secondary star fills its Roche lobe). 
In Fig. 1 (a, b, c) 
the mass--loss rate versus orbital period is presented for the following 
cases: (A)
sequence (I), $f_1$=$f_2$=0, (B) sequence (I), $f_1$=$f_2$=1, (C) sequence 
(II), $f_1$=$f_2$=1. At first, the mass 
exchange rate is rather high (proceeds on the thermal time--scale of the 
secondary star) and during this phase, the old neutron star will spin--up. 
Latter 
$\dot{M}$ decreases and near an orbital period of $\sim$ 2--4 hours, the mass 
accretion rate has its minimum value. Depending on the value of the surface 
magnetic field strength, the system may or may not be in the ``propeller 
phase''. How does the value of $\lambda$ influence the orbital evolution of 
the binary system? For example for sequence (A) and for   $\lambda$= 1.8 , 
the final orbital period is larger than the initial and for   $\lambda$ =1.0 
a short
period  binary system with low--mass helium white dwarf and millisecond 
pulsar is formed ($P_{orb}\sim$ 13 hours). Only for $\lambda$ $<$ 1 does 
the system evolve towards very short orbital periods (Fig.1).

\begin{figure}
%\epsfverbosetrue
\begin{center}
%\leavevmode
%\epsfxsize=7cm
\psfig{figure=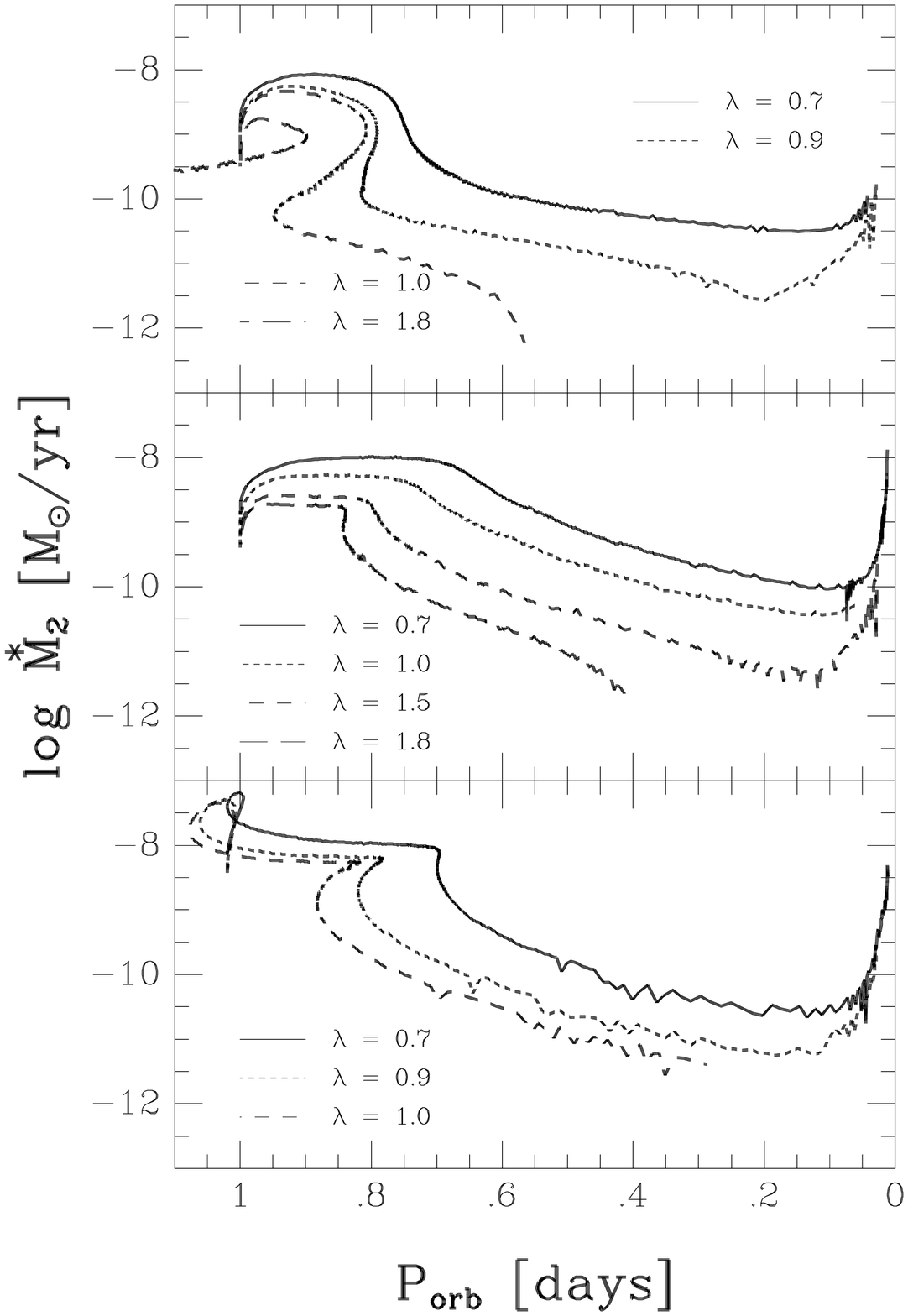,height=12cm,width=8cm}
\end{center}
\caption{The dependence of the mass loss rate as a function of orbital
period for three different evolutionary sequences (A) (upper panel), 
(B) (middle panel), (C) (lower panel). For details see
text.}
\end{figure}

\begin{figure}
%\epsfverbosetrue
\begin{center}
%\leavevmode
%\epsfxsize=7cm
\psfig{figure=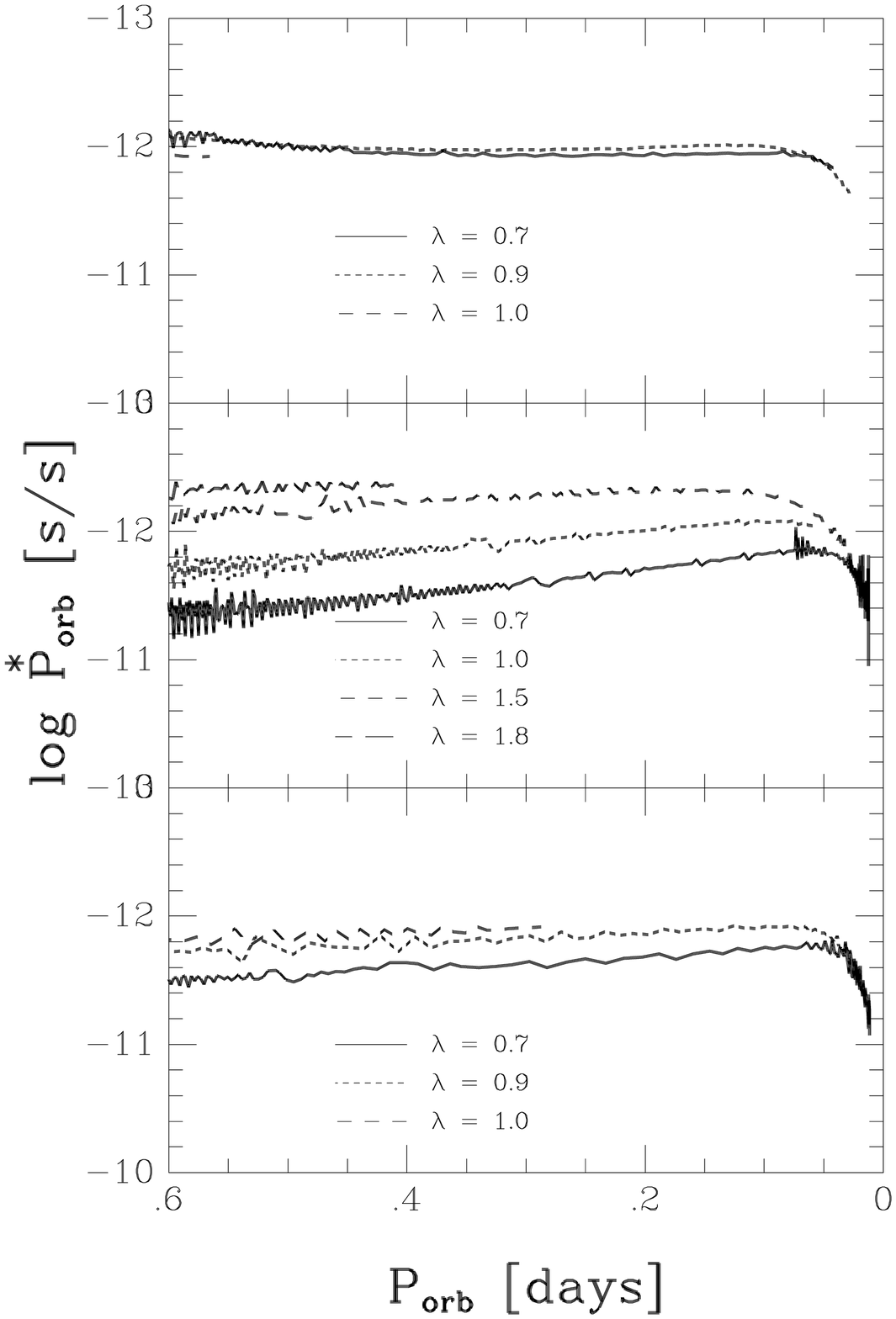,height=12cm,width=8cm}
\end{center}
\caption{The evolution of the orbital period derivative with the orbital 
period. Other details the same as for Fig. 1}
\end{figure}

In Fig. 2, a, b, c the variation of $\dot P_{orb} $  versus orbital period 
is shown
for the same cases as presented in Fig. 1 a, b, c. As is clearly seen in this 
Figure, orbital period changes  depend on 
the value of $\lambda$. 
In Table 1 we present $\dot{P}_{orb}$  and orbital period decay 
time--scale $\tau$=$P_{orb}/\dot{P_{orb}}$ values for various $\lambda$ when the orbital period is 
equal to 1.8 hours. The evolutionary time--scale obtained by Nice at al. 
(2000) is 200 Myr. From Table 1 we can see that a $\lambda$ close to 1  
agrees well with the observed value for PSR 1744--24A.The calculated mass of 
the secondary  
$\sim$ 0.15 $M_\odot$ near a  period 1.8 h also agrees well with the 
observed value (Lyne et al. 1990).

\section{ Conclusion}

If the observed orbital period decay is secular and
our suggestion is correct that the orbital evolution of PSR 1744--24A is 
driven by magnetic braking then this may be the first case when the orbital 
period decay by a magnetic braking mechanism has been measured.  According 
to our model calculations the best fit to the observed orbital 
period decay is obtained when $\lambda$ is near to one.  Which means that 
the value proposed by  Smith (1979)  $\lambda$=1.78 underestimates, that 
and proposed by Skumanich (1972) $\lambda$=0.73 sligtly,  overestimates  the 
magnetic braking efficiency. 

\section*{\sc Ackowledgements}
EE thanks Dr.B. Stappers for useful comments and in improving the text of 
the paper. Also we thank Dr.D.Nice for providing us results of observations before publication. Our special thanks anonymous referee for useful comments on the paper. EE acknowledges warm hospitality of the Astronomical Institue 
``Anton Pannekoek'' . This work was supported by NWO Spinoza grant 08--0 
to E.P.J.van den Heuvel and ESF grant N 4338.
At Warsaw, this work has been supported through grants 2--P03D--014--07 and 
2--P03D--005--16 by the Polish National Committee for Scientific Research.

\end{document}